\documentclass[sigconf,10pt,nonacm]{acmart}
\AtBeginDocument{%
  }

\setcopyright{acmcopyright}
\copyrightyear{2018}
\acmYear{2018}
\acmDOI{XXXXXXX.XXXXXXX}

\acmConference[Conference acronym 'XX]{Make sure to enter the correct
  conference title from your rights confirmation emai}{June 03--05,
  2018}{Woodstock, NY}
\acmPrice{15.00}
\acmISBN{978-1-4503-XXXX-X/18/06}

\usepackage{xcolor}
\usepackage{epstopdf}
\usepackage{graphicx}

\usepackage{xspace}
\newcommand{\edgelora}{\textsf{Edge2LoRa}\xspace}

\newcommand{\eed}{\textsf{E2ED}\xspace}
\newcommand{\egw}{\textsf{E2GW}\xspace}
\newcommand{\eas}{AS\xspace}

\begin{document}

\title{Enabling Edge processing on LoRaWAN architecture}


\author{Stefano Milani,\\ Ioannis Chatzigiannakis}
\affiliation{
  \institution{University of Roma "La Sapienza"}
  \city{Rome}
  \country{Italy} 
}
\email{stefano.milani@uniroma1.it}
\email{ichatz@diag.uniroma1.it}

\author{Domenico Garlisi}
\affiliation{
  \institution{University of Palermo}
  \city{Palermo} 
  \country{Italy} 
}
\email{domenico.garlisi@unipa.it}

\author{Matteo Di Fraia,\\ Patrizio Pisani}
\affiliation{%
  \institution{UNIDATA S.p.A.}
  \city{Rome} 
  \country{Italy} 
}
\email{m.difraia@unidata.it}
\email{p.pisani@unidata.it}

\renewcommand{\shortauthors}{Milani et al.}

\begin{abstract}
LoRaWAN is a wireless technology that enables high-density deployments of IoT devices. Designed for Low Power Wide Area Networks (LPWAN), LoRaWAN employs large cells to service a potentially extremely high number of devices. The technology enforces a centralized architecture, directing all data generated by the devices to a single network server for data processing. End-to-end encryption is used to guarantee the confidentiality and security of data. In this demo, we present \edgelora, a system architecture designed to incorporate edge processing in LoRaWAN without compromising security and confidentiality of data. \edgelora maintains backward compatibility and addresses scalability issues arising from handling large amounts of data sourced from a diverse range of devices. The demo provides evidence on the advantages in terms of reduced latency, lower network bandwidth requirements, higher scalability, and  improved security and privacy resulting from the application of the Edge processing paradigm to LoRaWAN.
\end{abstract}





\maketitle
\vspace{-4ex}
\section{Introduction}\label{sec:introduction}

In recent years, wireless technologies and mobile-generated traffic, including IoT, have rapidly expanded, becoming the largest segment of internet traffic. LoRaWAN, a widely adopted LPWAN technology, is an ideal solution for connecting various IoT devices with minimal infrastructure requirements~\cite{pagano-ieee}. In LoRaWAN the GateWays (GWs) have the role of simply forwarding all the traffic between the terminals, the End Devices (EDs), and the central Network Server (NS).
The traffic is then forwarded to the designated Application Server (AS) that securely handles, manages and interprets the application data. The Join Server (JS) is responsible for the activation of the ED~\cite{Alliance2017}. 
LoRaWAN supports two methods for registering and activating EDs on the network: i) Over-the-Air Activation (OTAA) and ii) Activation by Personalization (ABP), employing a secure communication protocol with encryption and authentication to ensure network security and privacy~\cite{Butun2018}. In ABP the session keys are pre-configured while in OTAA the session keys are generated during the ED activation phase. 

The centralised architecture and the provided activation methods of LoRaWAN forces the processing of the frames to be carried out exclusively in the cloud. Using the LoRa GWs as simple bridges places a lot of pressure on the central NS that needs to support a massive number of ED. This presents a substantial limitation on system performance in terms of scalability and can also impact time-sensitive IoT services.
%
%
%
\section{Edge2LoRa}
In this demo, we propose \edgelora a new LoRaWAN-based architecture to support edge processing that builds upon the OTAA~\cite{Milani2021}. \edgelora consists of several elements: the device registry, the gateway selection algorithm, the group key establishment method and the sensor data stream processing. Each ED is serviced by one \egw that carries out the data processing tasks on the received sensor data streams. 
\edgelora operates over the conventional LoRaWAN architecture, ensuring backward compatibility. Access to the data within the frames is facilitated by establishing a group key among the cloud (AS), edge (\egw), and end device (\eed).
Two shared session encryption keys are created between the \eed, \egw and AS: i) the Edge Session Encryption Key and the Edge Session Integrity Key. The former is used to enable secure encryption and decryption of the frame payload. The latter is used to check the integrity of the edge-specific frames. 
In this context, the proposed approach employs elliptic curve cryptography for generating cryptographic keys~\cite{Lauter2004}.

\section{Hardware set-up and software modules}
\label{sec:background}
\begin{figure}[t]
    \centering
    \includegraphics[width=1.0\linewidth]{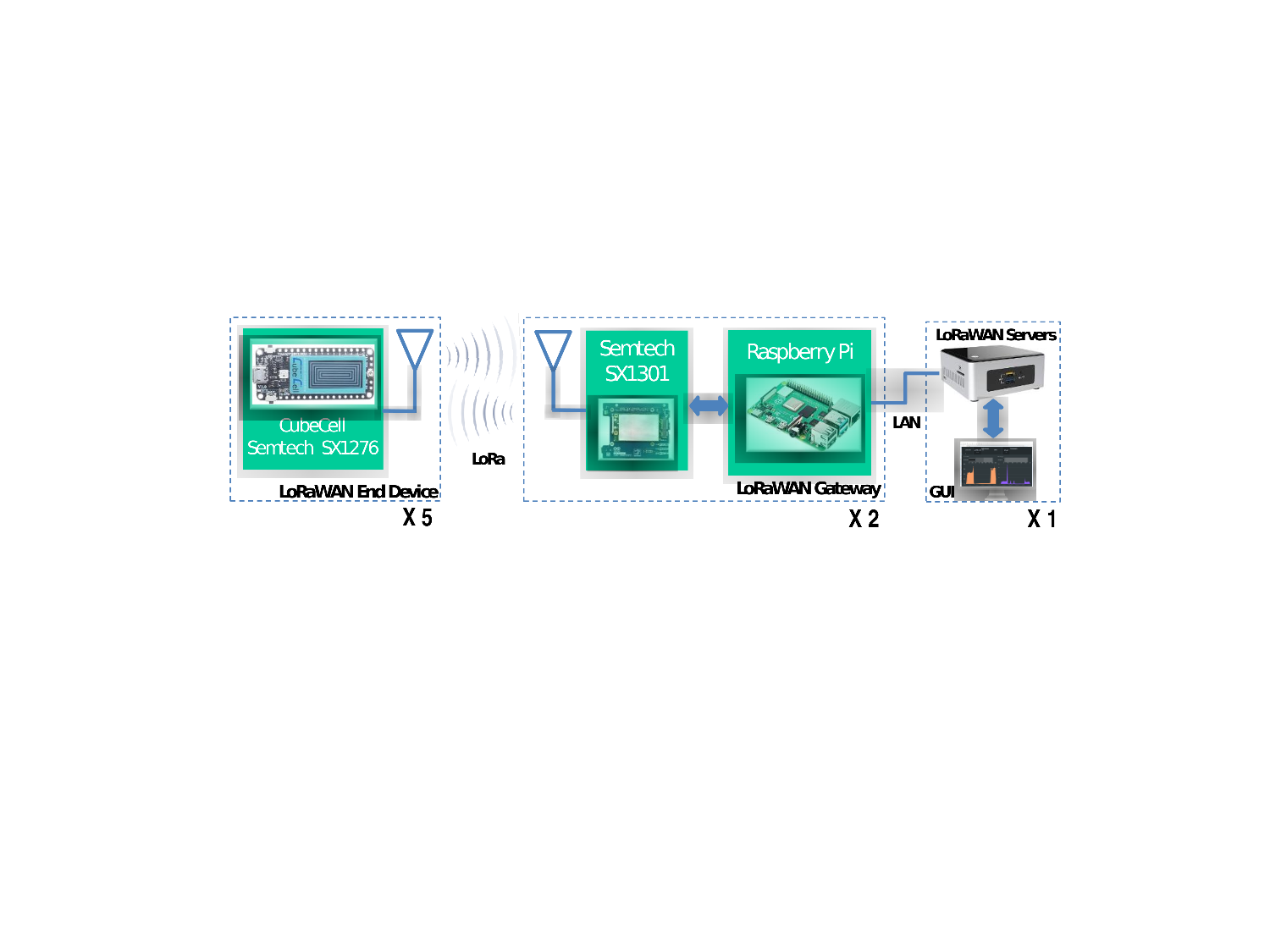}
    \caption{Demo hardware architecture.}
    \label{fig:setup}
    \vspace{-0.5cm}
\end{figure}
Fig.\ref{fig:setup} depicts the hardware architecture used for the demo, this includes 2 real {\em LoRaWAN GW}, implemented by using a Raspberry Pi 3 piloting the GW transceiver composed by a SEMTECH SX1301 chip, receive over-the-air the signal produced through 5 HTCC-AB01 “CubeCell“ series terminals, integrated the PSoC 4000 series MCU and SEMTECH SX1262 chip. Additionally, for each terminal, we connect a temperature/humidity/pressure sensor (BME280) to generate stream data. NS, AS and JS are deployed by an Intel NUC 5i5RYH Mini PC (8 GB of RAM). Additional monitor, keyboards and mouses (not shown in the figure), will be utilized for visualizing and controlling the demo dashboard.

\begin{figure}[t]
	\centering
	\includegraphics[width=1.0\linewidth]{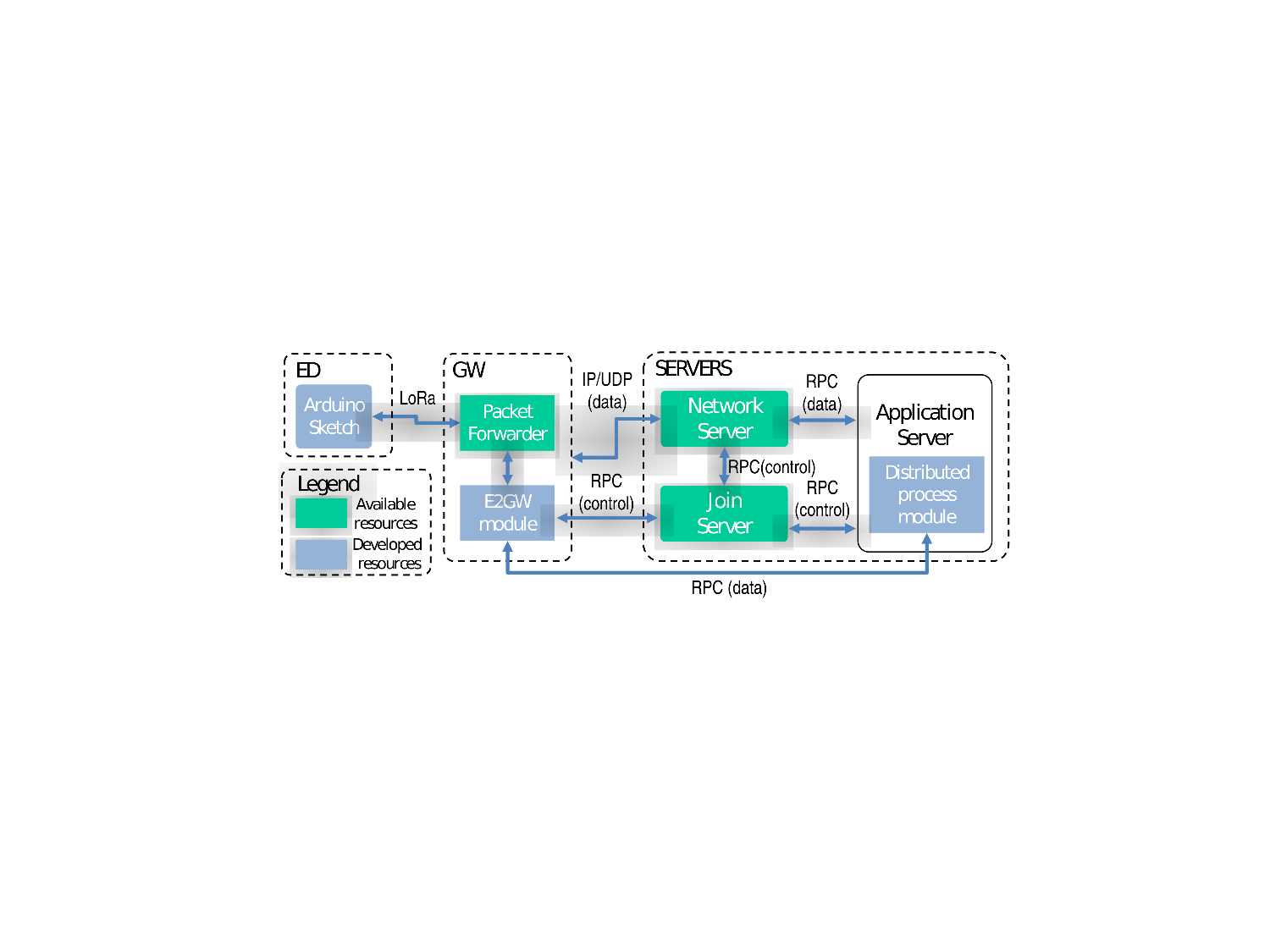}
	\caption{Edge2LoRa flow diagram.}
	\label{fig:swmodule}
	\vspace{-0.6cm}
\end{figure}

Fig.\ref{fig:swmodule} summarizes the software architecture of the demo. In the figure, according to the legend, two types of software modules are considered, the module already available as open source software and the module implemented to build the \edgelora architecture. 
It is important to highlight that the resulting system maintains backward compatibility, allowing seamless interoperability between legacy and new components, as demonstrated in the demo.
Three main blocks can be recognized in the figure: i) the ED including the Arduino Sketch module; ii) the GW including two modules, the Packet Forwarder (the legacy SEMTECH GW module) and the \egw module, iii) the SERVERS, including the NS and JS (both based on TheThingsStack), as well as the AS, which also hosts the demo Graphical User Interface (GUI). 
Finally, Fig.\ref{fig:swmodule} shows the selected protocols to interface the different modules.
\vspace{-0.4cm}
\section{Demonstration description}
\label{sec:background}
The objective of the demonstration is to provide insights on the \edgelora implementation in a real LoRaWAN network and show the performance gains under different scenarios, i.e. under different EDs configurations and data aggregation.
The 2 GWs are configured as legacy GW the first and as \egw the other. Legacy GW follows the classical data flow, while the \egw enables multiple aggregation functions and direct connection to the AS (exclusively for the traffic generated from the terminals configured as \eed).
The demo GUI enables two different control sections, the first has been designed to control the activity of the terminals and each terminal can be tuned as legacy or \eed. The terminal source rate and the message payload can be configured by the onboard button console.
The second GUI section receives the control of the applications in terms of aggregation function selection (including mean, sum, max and min of the sensor data) and window time. Moreover, the dashboard also tunes the available bandwidth of the link connections between GWs and AS, Indeed, the benefit of the aggregation is more evident when the link is slow.
%
%
Fig.\ref{fig:swmodule} also illustrates how the different elements' architectures are connected. For clarity, we describe only the uplink traffic. Data generated from the EDs follows the classical flow through the legacy GW, which forwards the data to the NS. Conversely, in the \egw, only frames from \eed activate the \egw module, where data stream operators apply transformations that are sent directly to the AS. The system uses many-to-one transformations, potentially aggregating data received from multiple frames. According to the radio coverage of the GWs and to the frames collision occurrence, only a subset of frames will be duplicated in the system.
For this reason, \edgelora relies on a duplicate detection filter (DDF) for identifying whether a given frame has previously appeared in a stream of data. The DDF is maintained by the \eas and identifies with no errors duplicate frames in constant time~\cite{geraud2020approaching}. 
The AS upon receiving a frame from the NS will assign a timeout before processing it. The timeout is set in a way such that it will allow the \egw to complete the processing of the operator. 
%
%

Finally, a visualization section is present in the dashboard to monitor system results, the number of frames received from the legacy path as well as from the \edgelora path and network statistics for: i) \textit{latency}: we show that by processing data closer to the source we reduce the latency associated with sending data to the cloud; ii) \textit{network traffic}: network traffic is significantly reduced by aggregating data at the Edge; iii) \textit{scalability}: in terms of generated data, here distributed computing resources can be easily scaled up or down based on the demand, without requiring additional infrastructure; iv) \textit{security and privacy capabilities}: configured message by EDs control can be visualized at the GW point, ciphered and in clear format. Data are processed at the Edge thus reducing the risk of large-scale data breaches and privacy violations.
In Tab.\ref{tab:results}, we present performance results in terms of latency and network traffic for a scenario with 2 EDs, a legacy device, and an \eed. We measure the average time difference between when the frames leave the Packet Forwarder module and when they reach the AS. The table compares the classical approach (Legacy column) to \edgelora. Notably, \edgelora enhances system performance by reducing latency from $955 \pm 4.6ms$ to $745 \pm 7.3ms$ (95\% Confidence Intervals) and data traffic from $120KB$ to $24KB$. These results were obtained during a period of activity where the EDs sent 100 frames, and the chosen aggregation function considered windows of 5 frames.
\begin{table}[t]
\centering
\begin{tabular}{lllll}
\cline{2-4}
\multicolumn{1}{l|}{} & \multicolumn{1}{l|}{Legacy} & \multicolumn{1}{l|}{\edgelora} & \multicolumn{1}{l|}{GAIN} &  \\ 
\cline{1-4}
\multicolumn{1}{|l|}{Latency} &
  \multicolumn{1}{l|}{\begin{tabular}[c]{@{}l@{}} $955 \pm 4.6 \ ms$ \end{tabular}} &
  \multicolumn{1}{l|}{\begin{tabular}[c]{@{}l@{}} $745 \pm 7.3 \ ms$\end{tabular}} & 
  \multicolumn{1}{l|}{\begin{tabular}[c]{@{}l@{}} $210\ ms$\end{tabular}} &
   \\ \cline{1-4}
\multicolumn{1}{|l|}{Data} &
  \multicolumn{1}{l|}{\begin{tabular}[c]{@{}l@{}} $120\ kBytes$\end{tabular}} &
  \multicolumn{1}{l|}{\begin{tabular}[c]{@{}l@{}} $24\ kBytes$\end{tabular}} & 
  \multicolumn{1}{l|}{\begin{tabular}[c]{@{}l@{}} $96\ kBytes$\end{tabular}} & 
   \\ \cline{1-4}
\end{tabular}
\caption{Comparison results between legacy and \edgelora in terms of latency and data traffic.}
\label{tab:results}
\vspace{-0.9cm}
\end{table}
%

\bibliographystyle{unsrt}
\bibliography{ref}

\end{document}